\documentclass[aps,prd,twocolumn,superscriptaddress,amsmath,amssymb,nofootinbib,floatfix]{revtex4-2}

\usepackage{graphicx}
\usepackage[colorlinks=true,citecolor=blue,linkcolor=blue,urlcolor=blue]{hyperref}

\newcommand{\rhoeff}{\rho_{\rm eff}}
\newcommand{\peff}{p_{\rm eff}}
\newcommand{\dd}{\mathrm{d}}
\newcommand{\OQ}{\Omega_{Q}}

\usepackage{scalerel}
\usepackage{tikz}
\usetikzlibrary{svg.path}
\usepackage[normalem]{ulem}

\newif\ifmarkup
\markuptrue
\definecolor{revblue}{HTML}{0047D6}
\definecolor{linkgreen}{HTML}{1E7B34}
\ifmarkup
  \newcommand{\add}[1]{\begingroup\color{revblue}\relax#1\endgroup}
  \newcommand{\del}[1]{\begingroup\color{revblue}\relax\sout{#1}\endgroup}
 
\definecolor{orcidlogocol}{HTML}{A6CE39}
\tikzset{
  orcidlogo/.pic={
    \fill[orcidlogocol] svg{M256,128c0,70.7-57.3,128-128,128C57.3,256,0,198.7,0,128C0,57.3,57.3,0,128,0C198.7,0,256,57.3,256,128z};
    \fill[white] svg{M86.3,186.2H70.9V79.1h15.4v48.4V186.2z}
                 svg{M108.9,79.1h41.6c39.6,0,57,28.3,57,53.6c0,27.5-21.5,53.6-56.8,53.6h-41.8V79.1z M124.3,172.4h24.5c34.9,0,42.9-26.5,42.9-39.7c0-21.5-13.7-39.7-43.7-39.7h-23.7V172.4z}
                 svg{M88.7,56.8c0,5.5-4.5,10.1-10.1,10.1c-5.6,0-10.1-4.6-10.1-10.1c0-5.6,4.5-10.1,10.1-10.1C84.2,46.7,88.7,51.3,88.7,56.8z};
  }
}

\newcommand\orcidicon[1]{\href{https://orcid.org/#1}{\mbox{\scalerel*{
\begin{tikzpicture}[yscale=-1,transform shape]
\pic{orcidlogo};
\end{tikzpicture}
}{|}}}}

\begin{document}

\title{The logotropic dark fluid as an energy diffusion in unimodular gravity}

\author{Miguel Cruz\orcidicon{0000-0003-3826-1321}}
\email{miguelcruz02@uv.mx}
\affiliation{Facultad de F\'{\i}sica, Universidad Veracruzana,
91097 Xalapa, Veracruz, M\'exico}

\author{Samuel Lepe\orcidicon{0000-0002-3464-8337}}
\email{samuel.lepe@pucv.cl}
\affiliation{Instituto de F\'{\i}sica, Pontificia Universidad Cat\'olica de
Valpara\'{\i}so, Casilla 4950, Valpara\'{\i}so, Chile}

\date{\today}

\begin{abstract}
We show that the logotropic dark fluid, proposed as a single-fluid unification of dark matter
and dark energy, is exactly dissipative unimodular gravity with pressureless matter and the
logarithmic energy diffusion function $Q(a)=3A\ln a+Q_{c}$. The correspondence is exact at the level of the homogeneous background, rather than asymptotic, and introduces no parameters beyond those already present in either
description. It provides a dynamical reinterpretation of the logotropic temperature $A$, whose origin was left open in the original proposal: up to the expansion
rate, $A$ is the rate at which energy is diffused from matter into the geometric sector,
$\dot Q=3AH$. The adiabatic sound speed of the effective fluid is
$c_{s}^{2}=A/\rho$, where $\rho$ is the unimodular matter energy density, and reaches the
speed of light at $a_{s}=a_{M}/2^{1/3}$; the matter energy density then vanishes and changes
sign at $a_{M}$, which is precisely the scale factor at which the model is known to become
phantom. Since $\rho$ coincides with the enthalpy density $\epsilon+P$ of the logotropic fluid, the phantom branch is the branch on which the matter sector carries
negative energy density and on which the apparent horizon entropy decreases. Finally, we show that a one-parameter saturating diffusion law removes these late-time pathologies while preserving the low-redshift phenomenology.
\end{abstract}

\maketitle

\section{Introduction}
\label{sec:intro}

Unimodular gravity (UG) is the simplest modification of general relativity (GR) that can
accommodate dissipative effects at the level of the field equations
\cite{Anderson1971,Unruh1989,Ellis2011}. Fixing the metric determinant replaces the Einstein
equations by their trace-free counterpart, whose Bianchi identity no longer enforces
$\nabla^{\nu}T_{\mu\nu}=0$ but instead promotes the cosmological term to an integration
constant sourced by a diffusion current,
\begin{equation}
J_{\mu}\equiv \nabla^{\nu}T_{\mu\nu},
\qquad
\Lambda=\Lambda_{0}+\int_{\ell}J ,
\label{eq:diffusion-current}
\end{equation}
so that in the conservative limit $J=0$ one recovers GR with a cosmological constant
identically \cite{Josset2017,Perez2019}. The resulting cosmologies have been studied
extensively, both with phenomenological diffusion functions
\cite{Corral2020,LinaresNucamendi2021,Landau2023,CruzCruzLepe2024} and, more recently, with a
diffusion law derived from a microscopic bath of quantum-gravity defects
\cite{Pellecchia2026}.

Independently of this description of gravity, Chavanis \cite{Chavanis2016} proposed the logotropic dark
fluid (LDF), a single-fluid unification of dark matter and dark energy obeying the logotropic
equation of state $P=A\ln(\rho_{r}/\rho_{P})$, with $\rho_{r}$ as the rest-mass density,
$\rho_{P}$ as the Planck density, and $A$ as a constant termed the logotropic temperature. The model is indistinguishable from $\Lambda$CDM at present, has no free parameters once $A$ is fixed by cosmological data, and accounts simultaneously for the observed universality of the surface density of dark matter haloes, the mass enclosed within $300$~pc of dwarf spheroidals, and the Tully-Fisher relation \cite{Chavanis2016}. Its weak point, acknowledged in the original paper, is that the logotropic equation of state is introduced heuristically, with only partial justifications sketched in relation to Cardassian models and to Tsallis generalized thermodynamics: {\it a justification from first principles is beyond the scope of this paper} \cite{Chavanis2016}.

The purpose of this note is to observe that the two frameworks describe the same background cosmology. We show in Sec. \ref{sec:correspondence} that the LDF is exactly dissipative UG with pressureless matter and a logarithmic diffusion function, and that the dictionary is fixed with no freedom. Section \ref{sec:consequences} draws the consequences we regard as significant: a dynamical interpretation of $A$ as a diffusion rate; the identification of the model's
phantom branch with a change of sign of the unimodular matter energy density, which we show to be unaffected by spatial curvature; and a saturating diffusion law that removes the late-time pathologies of the model.
Section \ref{sec:nottransfer} describes what \emph{does not} transfer, specifically the thermodynamic state of $A$.

We work with a spatially flat Friedmann-Lemaître-Robertson-Walker (FLRW) metric, set $8\pi G=c=1$, and use a barotropic equation of state $p=\omega\rho$ for the unimodular matter sector, with $\omega$ being a constant value. Overdots denote cosmic time derivatives. We normalize the present value of the scale factor to unity, $a_{0}=1$, so that a subscript zero denotes a present-day quantity.

\section{Dissipative unimodular gravity and its effective fluid}
\label{sec:frames}

For a spatially flat FLRW background, the unimodular field equations, together with the
integrated diffusion current \eqref{eq:diffusion-current}, read
\begin{align}
3H^{2}&= \rho+Q, \quad \dot H=-\tfrac12\left(\rho+p\right),
\label{eq:UG1}\\
& \dot\rho+3H(\rho+p)=-\dot Q ,
\label{eq:UG2}
\end{align}
where $Q$ collects the dynamical part of the cosmological term. From now on, we set $\Lambda_{0}=0$, which is no loss of generality since retaining it merely shifts
$Q$ by a constant. Note that the pressure of the fluid is not modified: we take, as usual, $p=\omega\rho$; the non-conservation enters as a source in the balance equation, not as a modification of the equation of state.

Adding the two equations in \eqref{eq:UG1} gives $\rhoeff+\peff=-2\dot H=\rho+p$ with
$\rhoeff\equiv3H^{2}$, so that
\begin{equation}
\rhoeff=\rho+Q,\qquad \peff=p-Q ,
\label{eq:effective}
\end{equation}
notice that according to our definitions \eqref{eq:UG2} yields
\begin{equation}
\dot\rhoeff+3H\left(\rhoeff+\peff\right)
=\left[\dot\rho+3H(\rho+p)\right]+\dot Q\equiv0 .
\label{eq:effcons}
\end{equation}
Dissipative UG is therefore exactly GR sourced by a \emph{conserved} total fluid: the
non-conservation characteristic of the theory is an internal partition of energy between
matter and geometry, not a property of the system as a whole. Equation \eqref{eq:effcons} is
an identity, not an assumption. Because the total effective fluid is a conserved perfect fluid, \eqref{eq:effcons} is equivalent to its adiabaticity: writing the Gibbs relation for a comoving volume $V\propto a^{3}$, $T\,\dd S=\dd(\rhoeff V)+\peff\,\dd V=V\!\left[\dot\rhoeff+3H(\rhoeff+\peff)\right]\dd t$, the identity \eqref{eq:effcons} gives $\dd S=0$. The non-conservation of the matter sector is thus an internal transfer of energy to the geometric sector at fixed total entropy, not a heat exchange with an external bath; this is the precise sense in which the model is adiabatic, in agreement with the LDF being an adiabatic fluid \cite{Chavanis2016}.

Since the effective fluid is a single
conserved fluid, it has a well-defined adiabatic sound speed, which follows from
\eqref{eq:UG2} alone:
\begin{equation}
c_{s}^{2}=\frac{\dd \peff}{\dd\rhoeff}=\omega+\frac{\dot Q}{3H\rho} .
\label{eq:cs}
\end{equation}
Moreover, the identity \eqref{eq:effcons} runs in both directions. Any single-fluid cosmology
can be read, at the background level, as a unimodular diffusion, with $Q$ and $\rho$ recovered from the total energy density and pressure through \eqref{eq:effective}; for $\omega=0$ this gives $Q=-\peff$ and $\rho=\rhoeff+\peff$, i.e., the unimodular matter density is the enthalpy density of the total fluid.

\section{The correspondence}
\label{sec:correspondence}

In the LDF model, the rest-mass density is conserved separately from the energy,
$\dot\rho_{r}+3H\rho_{r}=0$, so that $\rho_{r}=\rho_{r,0}a^{-3}$ holds exactly and the energy density splits as\footnote{According to \cite{Chavanis2016}, the relationship between energy density $\epsilon$ and rest-mass density $\rho$ is derived from the first law of thermodynamics, $d\epsilon = \frac{P+\epsilon}{\rho}d\rho$. Integrating yields $\epsilon = \rho c^2 + \rho \int \frac{P(\rho')}{\rho'^2} d\rho' = \rho c^2 + u(\rho)$, where $u(\rho)$ is the internal energy. Assuming a logotropic equation of state, $P = A \ln(\rho/\rho_P)$, we substitute $P$. Solving the integral by parts gives $u(\rho) = -A \ln(\rho/\rho_P) - A$. Thus, the total energy density is $\epsilon = \rho c^2 - A \ln(\rho/\rho_P) - A$, unifying a rest-mass term mimicking dark matter with an internal energy term mimicking dark energy.}
\begin{equation}
\epsilon=\rho_{r}+u(\rho_{r}) ,
\qquad
u=-P-A ,
\label{eq:LDFsplit}
\end{equation}
into a rest-mass term mimicking dark matter and an internal energy mimicking dark energy
\cite{Chavanis2016}. With the logotropic equation of state and $\rho_{r}\propto a^{-3}$ the
pressure is linear in $\ln a$,
\begin{equation}
P=A\ln\!\left(\frac{\rho_{r}}{\rho_{P}}\right)=-3A\ln a+{\rm const.}
\label{eq:Plog}
\end{equation}
We now take dissipative UG with pressureless matter, $\omega=0$, and the logarithmic diffusion function
\begin{equation}
Q(a)=3A\ln a+Q_{c},
\label{eq:Qlog}
\end{equation}
with $Q_{c}$ being a constant. The balance equation \eqref{eq:UG2} becomes $\dot\rho+3H\rho=-3AH$, which integrates in closed
form to
\begin{equation}
\rho(a)=\frac{\rho_{r,0}}{a^{3}}-A ,
\label{eq:rholog}
\end{equation}
where the integration constant is the present-day rest-mass density defined as $\rho_{r,0}\equiv \rho_{0}+A$ with $\rho_{0}=\rho(a=1)$. The effective fluid \eqref{eq:effective} is then
\begin{align}
\rhoeff&=\frac{\rho_{r,0}}{a^{3}}-A+3A\ln a+Q_{c} ,
\label{eq:efflog1}\\
\peff&=-3A\ln a-Q_{c} ,
\label{eq:efflog2}
\end{align}
which are precisely the energy density and pressure of the LDF, Eqs. \eqref{eq:LDFsplit} and
\eqref{eq:Plog}, under the identifications
\begin{equation}
Q=u+A ,\qquad \rho=\rho_{r}-A \add{.}
\label{eq:dictionary}
\end{equation}
The matching can be read off term by term. With $\rho_{r}=\rho_{r,0}a^{-3}$, the logotropic pressure \eqref{eq:Plog} is $P=-3A\ln a+A\ln(\rho_{r,0}/\rho_{P})$; comparing it with the effective pressure \eqref{eq:efflog2} fixes the integration constant of the diffusion function, $Q_{c}=-A\ln(\rho_{r,0}/\rho_{P})$, and yields $\peff=P$. The effective density then coincides with the total LDF energy density with no further freedom: using the dictionary $\rho=\rho_{r}-A$ and $Q=u+A$ one has $\rhoeff=\rho+Q=\rho_{r}+u=\epsilon$. Hence $(\rhoeff,\peff)=(\epsilon,P)$ at every value of $a$, so the unimodular effective fluid and the LDF are literally the same fluid at the level of the homogeneous background; the only nontrivial feature is that the constant $A$ ends up in the geometric sector, which shifts the matter density down to $\rho=\rho_{r}-A$. This is not a choice but is forced by $\omega=0$: with $p=0$, Eq.~\eqref{eq:effective} gives $\rho=\rhoeff+\peff$, so that
\begin{equation}
\rho=\epsilon+P=\rho_{r}-A,
\label{eq:enthalpy}
\end{equation}
i.e., the unimodular matter density is the enthalpy density of the LDF, as commented on before. The correspondence is exact, valid at all times rather than only asymptotically, and does not introduce any parameters beyond $A$ and $\rho_{r,0}$, which are already included in the LDF model. Its scope should nevertheless be stated precisely. Extending it to inhomogeneous perturbations requires specifying the diffusion current $J_{\mu}$ beyond the FLRW background, which neither description fixes. Throughout, $c_{s}^{2}$ denotes the adiabatic sound speed of the background fluid; it controls the propagation of perturbations provided these are adiabatic, $\delta\peff=c_{s}^{2}\,\delta\rhoeff$, as they are in the LDF, and our statements about superluminality are to be read in this sense.
Note the mild asymmetry in \eqref{eq:dictionary}: the constant $A$ is shuffled between the two sectors, so that the unimodular matter density is not exactly the rest-mass energy density but is shifted downwards by $A$. Since $A$ is small - $A/\epsilon_{0}=2.56\times10^{-3}$ for the values quoted in Ref. \cite{Chavanis2016} - the two agree to high accuracy at early times, and the difference matters only near the scale identified in Sec. \ref{sec:consequences}.

\section{Consequences}
\label{sec:consequences}

\subsection{The logotropic temperature is a diffusion rate}

Differentiating \eqref{eq:Qlog} gives
\begin{equation}
\dot Q=3AH ,
\label{eq:Qdotlog}
\end{equation}
so that $A$ is, up to the expansion rate, the rate at which energy is transferred from the
matter sector to the geometric one. For the case of constant $A$ we have that the diffusion
rate tracks the expansion rate, $\dot Q\propto H$, which is the simplest non-trivial closure
an energy diffusion law can have; the logotropic equation of state is what that closure looks
like when the same physics is written as a single-fluid equation of state.

This recasts the question left open in Ref.~\cite{Chavanis2016} in dynamical terms, although it does not by itself settle it: the heuristic equation of state is traded for an equally simple diffusion law, whose microscopic origin remains to be established. It also places the model
within a family. The energy diffusion functions most used in the unimodular literature are the barotropic law $Q=\alpha\rho$ \cite{Corral2020,CruzCruzLepe2024}, which gives a constant diffusion fraction $\OQ\equiv Q/(\rho+Q)$, and the continuous-spontaneous-localization law, in which the diffusion rate is set by the matter density, $\dot Q\propto\rho$ \cite{Corral2020,CruzCruzLepe2024}. The logarithmic law \eqref{eq:Qlog} also yields a running $\OQ$, but its rate is set by the expansion rate and is independent of $\rho$; as a consequence, the balance equation closes in terms of $a$ alone and integrates for any expansion history, which is what makes the model analytically solvable.

We also remark that in the expanding universe case, Eq.~\eqref{eq:Qdotlog} yields $\dot Q>0$ when $A>0$: energy flows from matter into the geometric sector, the same sign found in the microscopically derived law of Ref.~\cite{Pellecchia2026}. This might seem to clash with the second law, since for a fluid exchanging heat at a temperature $T$ positive entropy production requires $\dot Q<0$. There is no conflict: the total effective fluid is adiabatic, $\dd S=0$ [Eq.~\eqref{eq:effcons}], so $\dot Q>0$ does not describe heat released to a bath but the internal repartition of energy between the two sectors. Assigning a separate thermodynamic entropy (and hence a definite sign of its production) to the matter sector alone would require a temperature that the correspondence does not fix (Sec.~\ref{sec:nottransfer}). The entropy that is unambiguous here is the apparent-horizon entropy discussed below, whose production $\dot S_{h}\propto\rho$ is positive throughout the physical branch $a<a_{M}$.

\subsection{Two scales at which the model degrades}

Because the unimodular reading separates the matter density from the total, it attaches to the matter sector two features that, in the single-fluid description, appear only as properties of the enthalpy density $\epsilon+P$. The first is the sound speed. From \eqref{eq:cs} with $\omega=0$ and \eqref{eq:Qdotlog}, the adiabatic sound speed of the
effective fluid is
\begin{equation}
c_{s}^{2}=\frac{\dot Q}{3H\rho}=\frac{A}{\rho}=\frac{A}{\rho_{r,0}a^{-3}-A} ,
\label{eq:cslog}
\end{equation}
which is small whenever the matter density dominates and reaches the speed of light,
$c_{s}^{2}=1$, when $\rho=A$, i.e. at
\begin{equation}
a_{s}=\left(\frac{\rho_{r,0}}{2A}\right)^{1/3}=\frac{a_{M}}{2^{1/3}} ,
\label{eq:as}
\end{equation}
with $a_{M}$ defined as the quotient $(\rho_{r,0}/A)^{1/3}$. Reference \cite{Chavanis2016} computes the sound speed of its fluid independently, from $c_{s}^{2}=P'(\epsilon)$, and reports $c_{s}=c$ at $a_{s}=3.77$; using the same values and relations of \cite{Chavanis2016}, Eq. \eqref{eq:as} with $\Omega_{m,0}=0.274$ and $B\equiv A/\rho_{\Lambda}=3.53\times10^{-3}$, where $\rho_{r,0}=\Omega_{m,0}\,\epsilon_{0}$ and $\rho_{\Lambda}=(1-\Omega_{m,0})\,\epsilon_{0}$,
gives $a_{s}=3.767$. Since our calculation proceeds entirely through the unimodular variables
and \eqref{eq:cs}, the agreement is a consistency check of the correspondence
\eqref{eq:dictionary}; indeed, by Eq.~\eqref{eq:enthalpy}, Eq.~\eqref{eq:cslog} reads $c_{s}^{2}=A/(\epsilon+P)$, which is the LDF result written in the unimodular variables.

The second feature is the sign of the matter density. Eq. \eqref{eq:rholog} shows that the matter energy density vanishes and changes sign at
\begin{equation}
a_{M}=\left(\frac{\rho_{r,0}}{A}\right)^{1/3} ,
\label{eq:aM}
\end{equation}
which numerically is $a_{M}=4.746$ against the value $4.75$ at which
Ref. \cite{Chavanis2016} locates the minimum of $\epsilon(a)$ and the transition to phantom
behavior; the corresponding density, $\rho_{r}\text{\del{$c^{2}$}}/\epsilon_{0}=A/\epsilon_{0}
=2.56\times10^{-3}$, likewise matches the value quoted there. \emph{The phantom branch of the
logotropic model is the branch on which the unimodular matter sector carries negative energy
density.} By Eq.~\eqref{eq:enthalpy} this is equivalent to $\epsilon+P<0$, i.e., to the violation of the null energy condition by the LDF; what the unimodular description adds is not a new criterion but its physical reading in terms of the energy content of the matter sector.

That reading has an immediate consequence for horizon thermodynamics. For a spatially flat
FLRW universe, the apparent horizon sits at $r_{A}=1/H$ with entropy $S_{h}=\mathcal{A}_{h}/4G=8\pi^{2}r_{A}^{2}$
\cite{Bak2000,Cai2005,Wang2006}, where $\mathcal{A}_{h}=4\pi r_{A}^{2}$ and we used $8\pi G=1$, and \eqref{eq:UG1} gives
\begin{equation}
\dot S_{h}=-\frac{16\pi^{2}\dot H}{H^{3}}=\frac{8\pi^{2}(1+\omega)\rho}{H^{3}} ,
\label{eq:Shdot}
\end{equation}
so that in UG the sign of the horizon entropy production is controlled by the
\emph{bare} matter density alone. For $a>a_{M}$ one has $\rho<0$ and hence $\dot S_{h}<0$.
That phantom expansion is in tension with horizon thermodynamics, is of course known; what the correspondence adds is the identification of where the tension comes from in this particular model, namely the sign of the matter density, and the ordering of the two scales,
\begin{equation}
a_{s}=3.77\;<\;a_{M}=4.75 ,
\label{eq:ordering}
\end{equation}
so that the effective fluid becomes superluminal before it becomes phantom. Both lie far in
the future\del{,}\add{---}Ref.~\cite{Chavanis2016} estimates the phantom transition at some $25$~Gyr from
now\del{;}\add{---}and neither affects the agreement of the model with present data.

It is natural to ask whether spatial curvature could soften the two
features just described, since positive curvature admits effective phantom
behavior without violating the energy conditions \cite{CruzLepeCurvature}
and the unimodular equations extend straightforwardly to $k\neq0$
\cite{AguilarPerez2026}. It cannot, but the reason is instructive. Since we keep $a_{0}=1$, the curvature constant is dimensional, $k=-\Omega_{k,0}H_{0}^{2}$, and a closed geometry corresponds to $k>0$. The
balance equation \eqref{eq:UG2} is unchanged by curvature, so the unimodular matter
density is still $\rho=\rho_{r,0}/a^{3}-A$, the sound speed is still
$c_{s}^{2}=A/\rho$, and both $a_{M}$ and $a_{s}$ are independent of $k$.
What curvature does change is the phantom condition of the total
gravitating fluid: including the curvature contribution, $\rho_{\rm eff}=\rho+Q-3k/a^{2}$ and $p_{\rm eff}=p-Q+k/a^{2}$, the combination $\rho_{\rm eff}+p_{\rm eff}=-2\dot H=(\rho+p)-2k/a^{2}=\rho-2k/a^{2}$ for $p=0$; so
super-acceleration now sets in at $\rho=2k/a^{2}$ rather than at $\rho=0$.
For a closed geometry ($k>0$) this threshold lies slightly below $a_{M}$,
opening a window $a_{\rm ph}<a<a_{M}$ in which the total fluid is already
phantom while the matter density is still positive; this is the case of a curvature-driven,
DEC-preserving acceleration of the kind discussed in
\cite{CruzLepeCurvature}. Since the
curvature scale $2k/a^{2}$ is far below $A$ for any observationally
admissible curvature (for $|\Omega_{k,0}|\sim10^{-3}$ one has $2k/a_{M}^{2}\sim10^{-2}A$), the phantom threshold shifts only marginally and the
three scales keep the ordering $a_{s}<a_{\rm ph}<a_{M}$, so the effective
fluid still turns superluminal before it turns phantom, exactly as in the
flat case; curvature merely narrows the interval between the two. Beyond $a_{M}$
the matter density turns negative and drives both the total phantom and the
local pathologies together. That the pathologies track $\rho$ and not the
total is most transparent for the horizon entropy: the curved apparent
horizon obeys $H^{2}+k/a^{2}=(\rho+Q)/3$, so the curvature cancels
identically, $S_{h}=24\pi^{2}/(\rho+Q)$ and $\dot S_{h}=72\pi^{2}H\rho/(\rho+Q)^{2}$, where the continuity equation was used, and $\dot S_{h}<0$ still onsets
exactly at $a_{M}$. Nor is there any escape through a turning point: the
curvature term decays as $a^{-2}$ while the effective density grows,
$\rho+Q\to3A\ln a$, so $H^{2}$ stays positive through $a_{M}$ and,
asymptotically, $\rho+p-2k/a^{2}\to-A$ independently of $k$. The phantom
transition is thus intrinsic to the matter-geometry partition: curvature
can relabel where the total fluid first violates the null energy condition,
but it leaves the negative-density branch—and with it the superluminality
and the entropy decrease—exactly where the flat analysis places them. The
identification of the phantom branch with $\rho<0$ is, strictly, a statement
about the spatially flat case.

\subsection{A saturating diffusion removes the pathologies}
\label{sec:saturating}

The three features localized above\del{:}\add{---}the change of sign of $\rho$, the
superluminal sound speed, and the decreasing horizon entropy\add{---}share a single
origin. The logarithmic diffusion \eqref{eq:Qlog} never switches off: with
$A$ constant the source drains the matter sector at the fixed rate
$\dot Q=3AH$, so the finite reservoir $\rho_{r,0}/a^{3}$ is eventually depleted and
$\rho=\rho_{r,0}/a^{3}-A$ overshoots into negative values at $a_{M}$. Since curvature
does not enter the balance equation, a possibility able to cure this is the
diffusion law itself. A natural criterion is that the transfer rate saturate
once the matter density has been sufficiently diluted, which corresponds to the
behavior of the microscopically derived law of Ref. \cite{Pellecchia2026},
whose current scales as $a^{-3}(\rho+p)$ and whose effective cosmological
term freezes to a constant; since that law was derived for the primordial universe, the analogy is structural rather than quantitative. Consider therefore the one-parameter closure
\begin{equation}
\dot Q=\frac{3AH}{\left[1+(a/a_{\ast})^{3}\right]^{2}},
\label{eq:Qsat}
\end{equation}
which reduces to the logotropic rate $\dot Q\to3AH$ for $a\ll a_{\ast}$ and
falls as $\dot Q\to3AH\,(a_{\ast}/a)^{6}$ for $a\gg a_{\ast}$; in the de
Sitter regime reached at late times (see below), and for $a_{\ast}<a_{M}$, this is $\dot Q\propto a^{-3}(\rho+p)$, the same scaling as the current of Ref.~\cite{Pellecchia2026}. With $\omega=0$ the
balance equation \eqref{eq:UG2} again integrates in closed form: with $x\equiv(a/a_{\ast})^{3}$, multiplying \eqref{eq:UG2} by $a^{3}$ gives $\dd(a^{3}\rho)/\dd a=-3Aa^{2}/(1+x)^{2}$, which integrates elementarily to
\begin{equation}
\rho(a)=\frac{\rho_{r,0}}{a^{3}}-\frac{A}{1+(a/a_{\ast})^{3}}.
\label{eq:rhosat}
\end{equation}
where $\rho_{r,0}$ now denotes the integration constant, related to the present matter density by $\rho_{r,0}=\rho_{0}+A/(1+a_{\ast}^{-3})$, which reduces to the logotropic definition for $a_{\ast}\gg1$; we keep $a_{M}\equiv(\rho_{r,0}/A)^{1/3}$. Indeed, $\dot\rho+3H\rho=3AH\!\left[x/(1+x)^{2}-1/(1+x)\right]=-3AH/(1+x)^{2}=-\dot Q$, as required.
The drained energy
$\int a^{3}\,\dd Q=Aa^{3}/[1+(a/a_{\ast})^{3}]$ now saturates to the constant
$Aa_{\ast}^{3}$ instead of growing without bound. For $a\ll a_{\ast}$ both
\eqref{eq:Qsat} and \eqref{eq:rhosat} coincide with the logotropic
expressions \eqref{eq:Qlog}-\eqref{eq:rholog}, so the sound speed
\eqref{eq:cslog}, the scale $a_{s}$, and the full low-redshift phenomenology
are reproduced; because $A/\epsilon_{0}=2.56\times10^{-3}$ is tiny, the
correction to the present-day dynamics is negligible for any $a_{\ast}\gtrsim1$.

The single new parameter is the saturation scale $a_{\ast}$, and the whole
cure follows from one inequality. The numerator of \eqref{eq:rhosat} is
$A\big[a_{M}^{3}+a^{3}(a_{M}^{3}/a_{\ast}^{3}-1)\big]$, so
\begin{equation}
\rho(a)>0\ \text{for all }a\quad\Longleftrightarrow\quad a_{\ast}\add{\le}a_{M},
\label{eq:poscond}
\end{equation}
and provided \eqref{eq:poscond} holds the matter density remains positive at
all times, decaying as $\rho\to(\rho_{r,0}-Aa_{\ast}^{3})/a^{3}$ for $a_{\ast}<a_{M}$; in the marginal case $a_{\ast}=a_{M}$ this leading term cancels and $\rho=Aa_{M}^{6}/[a^{3}(a^{3}+a_{M}^{3})]\propto a^{-6}$. The sound speed,
\begin{equation}
c_{s}^{2}=\frac{\dot Q}{3H\rho}
=\frac{A}{\rho\,\left[1+(a/a_{\ast})^{3}\right]^{2}},
\end{equation}
carries the cutoff factor exactly where the logotropic model became
superluminal, and stays subluminal, $c_{s}^{2}<1$, throughout the evolution
for every $a_{\ast}\le a_{M}$; indeed $c_{s}^{2}<1$ is equivalent to $\rho\,[1+(a/a_{\ast})^{3}]^{2}>A$, which holds for all $a$ whenever $a_{\ast}\le a_{M}$. For $a_{\ast}<a_{M}$ one has $c_{s}^{2}\to0$ in the future, whereas in the marginal case $c_{s}^{2}=[1+(a_{M}/a)^{3}]^{-1}\to1^{-}$, and the late-time scaling $\dot Q\propto a^{-3}\rho$ is also lost. We therefore take $a_{\ast}<a_{M}$ strictly in what follows. The horizon entropy obeys $\dot S_{h}\propto\rho>0$ and therefore increases at all times. Finally, integrating \eqref{eq:Qsat} gives $Q=Q_{\infty}+A\!\left[\ln\frac{x}{1+x}+\frac{1}{1+x}\right]$, which tends to the constant $Q_{\infty}$. Therefore, $\rho_{\rm eff}=\rho+Q\to Q_{\infty}$ and the
universe tends toward de Sitter; since $\rhoeff+\peff=\rho>0$ holds throughout, the null
energy condition is never violated, and $w_{\rm eff}\to-1^{+}$ from the
quintessence side, meaning the model never becomes phantom.

The price is a single additional parameter, the saturation scale $a_{\ast}$,
constrained only by \eqref{eq:poscond} and otherwise free: since every
departure from the logotropic law occurs at $a>a_{\ast}\gtrsim1$, it is
unconstrained by present data and the model remains indistinguishable from
$\Lambda$CDM today, exactly as the original proposal. What \eqref{eq:Qsat}
shows is that the late-time pathologies are not intrinsic to the unimodular
reading but to the specific unbounded growth of the logarithmic diffusion:
a closure that interpolates between the logotropic rate $\dot Q\propto H$ at
low redshift and a matter-tracking, self-limiting rate at large $a$ retains
the successful phenomenology while curing $\rho<0$, the superluminality, and
the entropy decrease at once.

\section{What does not transfer}
\label{sec:nottransfer}

One point deserves emphasis, since the word {\it temperature} appears on both sides of the
dictionary with different meanings. The logotropic temperature $A$ is not a thermodynamic
temperature. Reference \cite{Chavanis2016} obtains it as the coefficient of an equation of
state and interprets it, following Refs.~\cite{McLaughlin1996,ChavanisSire2007}, as a
generalized temperature conjugate to the log-entropy $S_{L}=\int\ln\rho\,\dd\mathbf{r}$ within generalized thermodynamics \cite{Tsallis2009}; in that sense the model describes an
``isothermal'' universe, but the isothermality refers to a logotropic rather than a linear
equation of state. The model itself is adiabatic, $\dd S=0$.

Consequently the correspondence \eqref{eq:dictionary} says nothing about the thermodynamic
temperature of the unimodular matter sector, which must be obtained from the
Gibbs equation and the integrability condition for the entropy \cite{Maartens1996}. What does
transfer is the kinematic decomposition \eqref{eq:LDFsplit}, with the rest-mass density
separately conserved, which is what makes the dictionary possible in the first place. We note
that this decomposition sits particularly naturally in UG: the diffusion
current \eqref{eq:diffusion-current} violates $\nabla^{\nu}T_{\mu\nu}=0$ but need not violate
$\nabla_{\mu}N^{\mu}=0$, so that particle number can be conserved while the energy per particle
changes. Developing that observation into a temperature law is left for future work.

\section{Concluding remarks}
\label{sec:conclusions}

At the level of the homogeneous background, the logotropic dark fluid is dissipative unimodular gravity with pressureless matter and the
diffusion function $Q=3A\ln a+Q_{c}$. The correspondence is exact and parameter-free, and it
supplies a dynamical reading of the logotropic temperature as a diffusion rate,
$\dot Q=3AH$, the simplest closure available to an energy diffusion law. This recasts, without by itself settling, the question of principle left open in Ref. \cite{Chavanis2016}: the heuristic equation of state is traded for a diffusion law whose microscopic origin is still to be found.

The reading also attaches two features of the model to the matter sector, whose energy density is the enthalpy density $\epsilon+P$ of the LDF. The effective sound speed
$c_{s}^{2}=A/\rho$ reaches unity at $a_{s}=a_{M}/2^{1/3}$, reproducing through the unimodular
variables a result obtained in Ref.~\cite{Chavanis2016} by a different route; and the phantom
transition at $a_{M}$ is the point at which the unimodular matter density changes sign (equivalently, where $\epsilon+P$ vanishes), which is also where the apparent-horizon entropy begins to decrease.

More broadly, the identity \eqref{eq:effcons} means that the reading performed here is
available for any single-fluid cosmology: each such model is, at the background level, a unimodular diffusion, with $Q$ read off from the total pressure. Whether that reading is illuminating depends on whether the resulting $Q$ is simple, and in the logotropic case it is as simple as it could be. Applying the same procedure to other unified dark sector models, and asking which of them correspond to diffusion functions admitting an independent motivation, seems to us a question worth pursuing.

Finally, the same reading suggests how the model's late-time behavior might
be repaired. The pathologies localized above are not intrinsic to the
unimodular description but to the unbounded growth of the logarithmic
diffusion \eqref{eq:Qlog}, which drains the matter sector past zero; a
closure that interpolates between the logotropic rate $\dot Q\propto H$ at
low redshift and a self-limiting, matter-tracking rate at large $a$ - so that
the diffusion saturates before the reservoir is exhausted - keeps the matter
density positive, the sound speed subluminal and the horizon entropy
increasing, and yields a non-phantom de Sitter future, all at the cost of a
single scale $a_{\ast}<a_{M}$ that is unconstrained by present data. Whether such a saturating law arises from an
underlying diffusion mechanism, as in the microscopically derived current of
Ref. \cite{Pellecchia2026}, rather than being imposed by hand, is to us the
most concrete question left open by the correspondence.

\begin{acknowledgments}
MC work was partially supported by S.N.I.I. (SECIHTI--M\'exico). SL acknowledges the FONDECYT
grant N$^{\circ}$~1250969, Chile.
\end{acknowledgments}

\end{document}